# Reliable Graph-based Collaborative Ranking

Bita Shams [a] and Saman Haratizadeh [a]
[a] Faculty of New Sciences and Technologies, University of Tehran, Tehran, Iran .

**Abstract**

GRank is a recent graph-based recommendation approach the uses a novel heterogeneous information network to model users' priorities and analyze it to directly infer a recommendation list. Unfortunately, GRank neglects the semantics behind different types of paths in the network and during the process, it may use unreliable paths that are inconsistent with the general idea of similarity in neighborhood collaborative ranking. That negligence undermines the reliability of the recommendation list generated by GRank.

This paper seeks to present a novel framework for reliable graph-based collaborative ranking, called ReGRank, that ranks items based on reliable recommendation paths that are in harmony with the semantics behind different approaches in neighborhood collaborative ranking. To our knowledge, ReGRank is the first unified framework for neighborhood collaborative ranking that in addition to traditional user-based collaborative ranking, can also be adapted for preference-based and representative-based collaborative ranking as well. Experimental results show that ReGRank significantly improves the state-of-the art neighborhood and graph-based collaborative ranking algorithms.

## 1. Introduction

Recommender systems are sort of information filtering systems that help people through filtering the items that are in interest of the users. Although recommender systems can be designed to make recommendations based on distributed knowledge in peer-to-peer architectures [4,6], here we focus on the standard centralized recommendation service architecture that analyzes a centralized knowledge base for making recommendations.

Collaborative filtering is the main class of recommendation algorithms which exploits users' historical feedbacks to predict and recommend what users will require more in the future. Traditional collaborative filtering algorithms, collaborative rating methods, generate a model that correctly predict how users will rate the items. Recently, collaborative filtering algorithms are directed to learn the users' ranking over items. The latter approach, called collaborative ranking, has gained more attention as recommendation algorithms should improve quality of Top-N recommendation that is inherently a ranking task[13,26].

Neighborhood collaborative ranking algorithms, also called NC-Rank methods, typically calculate users' similarities through Kendall correlation and its variants that take into account the number of agreements/disagreements of users over common pairwise comparisons of items, and find the most similar users to the target user. Then they analyze the pairwise comparisons of those similar users to predict the ranking of items for the target user [1,13,18,30]. NC-Rank algorithms generally get in trouble in sparse datasets where information is not rich enough to calculate reliable similarities and to find useful neighborhoods for estimating the preferences of users. More clearly, in sparse datasets users rarely have common pairwise preferences and so, most of estimated similarity values, obtained by Kendall and its variants[12,30], will be equal to zero[18]. Even if there exists a set of common pairwise preferences between two users, it is not large enough to ensure that similarity values are reliable [18,19]. In such a situations, NC-Rank algorithms form the neighborhood randomly and so, they fail to infer the true ranking of items for the target user [18]. Graph-based collaborative ranking (GRank) [19] resolves this sparsity issue by calculating extended similarities among entities in a heterogeneous information network, called Tripartite preference graph (TPG) that reflects priorities of users over items. TPG reflects the relations between entities of neighborhood collaborative ranking that are users, pairwise comparisons and items' representatives (that refer to the winner or loser sides of items). It uses the personalized PageRank algorithm on the graph to propagate the rank from a target node to other nodes in the graph via existing paths in the network. The amount of rank propagated from a target user node to a an item node is then used to estimate the closeness of that item to the target user, which in turn determines the position of that item in the sorted recommendation list of items for that user.

Heterogeneous information networks, such as TPG, are comprised of different types of entities resulting in different types of paths each reflecting a specific semantic. For instance, from the rating-oriented perspective, a certain type of path may connect the target user to items that are similar to his favorite items, while another kind of path connects a user to items that his friends like them. These types of relations are in harmony with different classes of neighborhood collaborative filtering; the former one follows the idea of item-based recommendation approach, while, the latter one follows the idea of user-based recommendation approach. Therefore, it is important for a graph-based recommendation algorithm to take into account the semantics behind different types of paths in heterogeneous networks [8,9,27]. Otherwise, it might score items based on unreliable paths that are inconsistent with the semantic behind the desired approach of neighborhood recommendation.

Unfortunately, GRank neglects these semantics and interprets each existing type of path between any two nodes of the graph, as an evidences for closeness of those nodes. Therefore, it needs to be investigated whether GRank scores items through reliable paths (i.e. paths that are in harmony with the desired concept of similarity) or not, and if it does not, then we need a way to identify the reliable paths in TPG, before using them for making recommendations.

To handle these issues we first formalize different perspectives to neighborhood collaborative ranking, that are user-based, preference-based, and items' representative-based collaborative ranking. Thereafter, we determine which paths of TPG are in harmony with each perspective, and call them reliable recommendation meta-paths for that perspective. Once the reliable recommendation meta-paths are extracted, we show that GRank score items through paths that are not reliable in any classes of neighborhood collaborative ranking. Finally, we present a framework for graph-based collaborative ranking, called ReGRank, which guarantees to score items only based on reliable paths. ReGRank is adapted to follow the semantic behinds user-based, pairwise preference-based and representative-based neighborhood collaborative ranking.

The main contributions of this paper can be summarized as follows:

- We present a novel approach to describe a large set of paths on heterogeneous information with a short string, called *string description* of meta-paths. That enables us to define and analyze reliable recommendation paths from different perspectives in neighbor-based collaborative ranking.
- We formalize the user-based, representative-based, and preference-based recommendation approaches in a graph-based collaborative ranking framework called ReGRank. To our knowledge, this is the first comprehensive research on different classes of neighbor-based collaborative ranking algorithms.
- We systematically show how each approach in neighborhood collaborative ranking can be modeled through a particular set of paths in TPG. Also, we show that TPG contains some paths that are not reliable for calculating similarity in any class of NC-Rank methods.
- We introduce three novel network structures $g_{UNC}, g_{PNC}$, and $g_{RNC}$ whose all paths are reliable from different perspectives of NC-Rank and we show how to construct and use those network structures to make reliable recommendations in ReGRank.
- We provide a comprehensive set of experiments to evaluate different classes of ReGRank. The results show that user-based and representative-based ReGRank improve current state-of-the art neighborhood and graph-based collaborative ranking methods.

The rest of this paper is organized as follows. Section 2, we discuss the current collaborative ranking methods especially those that follows neighborhood recommendation approach which covers the scope of this paper. Then, we summarize the preliminaries in graph-based collaborative ranking in section 3. Next, in section 4, we show how to briefly describe recommendation meta-paths of recommendation algorithms by a string descriptions. The concept of string description enable us to formally define reliable recommendation meta-paths of neighborhood collaborative ranking in section 5. Thereafter, we present a novel framework for reliable graph-based collaborative ranking in section 6. Finally, we present experimental evaluations of our approach in section 7, and conclude the whole paper with directions to future works in section 8.

## 2. Related work

As the scope of this paper lies in the category of neighborhood collaborative ranking here we will briefly review the alternative approach to collaborative ranking, called latent factor model, and then we will discuss the neighborhood collaborative ranking algorithms in more details. We also mention that current neighborhood collaborative ranking responds to the queries using a client/server architecture in which there is a centralized data center that encompass all available information.

## 2.1. Latent factor models

Latent factor model (LFM) approach of collaborative ranking transforms users and items to a latent common feature space in which we can correctly estimate interest of users to items through calculating inner product of their corresponding feature vectors. Cofi-Rank-NDCG [31] directly optimize the convex upper bound of a ranking loss function, called normalized discounted cumulative gain NDCG. Shi. et al [25] presented a LFM approach that lead to correct prediction of top-1 probabilities of each item for user. URM[34] aggregates ranking and rating-oriented loss functions to improve the quality of total ranking. ListPMF[11] adapts probabilistic matrix factorization approach to maximize the log- posterior of the predicted and observed preference orders of users. BoostMF[2] aggregates boosting and matrix factorization models to improve the quality of top-N recommendation. Another group of LFM algorithms maps users and items to a latent feature space which results in correct prediction of users' pairwise preferences. CofiRank-ordinal is one of pioneers in this category which minimizes the number of dis-concordant comparisons among items[32] when rating data is available. Bayesian personalized Ranking (BPR) is a pairwise collaborative ranking framework for users' implicit feedbacks. ABPR[17] adapts Bayesian personalized ranking for heterogeneous implicit feedbacks. CofiSet[16] learns the pairwise preferences among item sets rather than single items. PushAtTop [3] is another LFM approach that weighs the pairwise comparisons corresponding to items that are highly preferred by each user. CLIMF[23] and xCLIMF[24] optimizes the mean reciprocal rank (MRR) of the recommendation list. UOCCF[10] is one of the recent collaborative ranking framework which aggregates CLIMF and probabilistic matrix factorization to improve top-n recommendation quality in case of implicit feedbacks.

We note that this paper focus on neighborhood collaborative ranking, and so, our algorithm is totally different from LFM techniques as it does not map users and items to any other feature space. Instead, it explores the concept of similarities between entities like users and items to estimate a target user's priorities.

## 2.2. Neighborhood methods

Neighborhood class of collaborative ranking methods, which covers the current research, explores opinions of similar users to predict the total ranking of items for the target user. Although some researchers have investigated list-wise[29] or point-wise[7] approach to calculate users' similarity, most of current NC-Rank algorithms follow the pairwise approach in which takes into account users' pairwise preferences over items[1,12,14,30,33]. More clearly, most of current NC-Rank algorithms follow a three step framework. The first step is to calculate users' similarities according to their agreement/disagreement over common pairwise preferences. The second step, is to estimate a preference matrix for the target user based on the pairwise preferences of k-most similar users to him. Finally, the third step is to aggregate the estimated pairwise preferences to infer the total ranking of items for the target user. This framework is originally introduced by EigenRank [13] which uses Kendall correlation to estimate users' similarities, and a random walk approach for ranking inference. Wang et al. [30] has slightly modified Kendall correlation in order to weigh different pairwise preferences according to their popularity and their strength in a rating dataset. Kalloori et al.[5] have suggested to calculate the similarities using EDRC metric, that models each user $u$ as a preference graph of items whose link $e_{ij}$ indicates the preference of item $i$ over item $j$ for user $u$. Then, it calculates the similarities between preferences graphs of users to estimate the similarities of users.

The main flaw of the mentioned algorithms is their inefficiency in sparse data when users rarely have common pairwise preferences. SibRank[18] exploits a signed bipartite network to calculate extended similarities even among users that do not have any common pairwise preferences. Although SibRank is able to calculate users' similarities in sparse data, it can't estimate the preferences for the target user when no information is available in his neighborhood. GRank[19] is another network-oriented collaborative ranking framework which applies personalized PageRank over the so called Tripartite Preference Graph (TPG) to directly estimate the interest of users to the items. This approach enables GRank to assign higher scores to items that are popular in the neighborhood while it is still capable of using information that is available outside of the neighborhood. However, GRank only takes into account the length and number of existing paths between a target user and item representatives without considering the semantics behind different types of paths in TPG. This ignorance affects GRank's reliability as it might score items based on the paths that do not reflect a reasonable sense of relevance between a target user and an item representative, or even are in contrast to the concept of similarity in the desired neighborhood recommendation approach.

Our suggested framework, called ReGRank, is different from current NC-Rank algorithms from different perspectives. First, it does not rely on common pairwise comparisons to calculate users' similarities as in [13] [30]. Instead, it exploits a graph-based proximity measure that is able to calculate transitive similarities even when users have no common pairwise preferences. Moreover, ReGRank directly estimates interests of users to items base on all available information in the system while most of NC-Rank algorithms define a neighborhood around the target user and only explore opinion of users in that neighborhood for scoring items[1,13,18,30,33]. Also it should be noted that unlike GRank, ReGRank takes into account the semantics behind each type of path in TPG and only uses reliable paths which follow the concept of similarity in a neighborhood collaborative ranking approach. Finally, note that while traditional NC-Rank algorithms lie in the category of user-based collaborative ranking methods [1,13,18,30,33], GRank[19] can't be placed in any category, not user based nor item based methods. On the other hand, ReGRank is a unified framework that can be adapted for user-based, preference-based and representative-based neighborhood collaborative ranking as well.

## 3. Preliminaries

A rating-oriented system can be modeled as a bipartite network with adjacency matrix B whose element $B_{i,j}$ indicates the opinion of a user $u_i$ to an item $i_j$. However, this approach is not applicable in ranking-oriented systems in which users' feedbacks to items are not absolute values and the choices of the users should be analyzed with regard to the choice context. More clearly, ranking-oriented data set can be represented in the form of set $O = \{o = <u, i, j>, u \in U, i \in I \text{ and } j \in I\}$ where $o = <u, i, j>$ denotes that user $u$ has preferred item $i$ over item $j$. Such a dataset might contain a record indicating that user $u$ prefers item $A$ over item $B$ while some other item like $C$ may be preferred over $A$ by user $u$. That means that there are two sides for each item $i$, the desirable side, $i_d$, that refers to when a user prefers $i$ over another item, and the undesirable side, $i_u$, that refers to when a user prefers another item to $i$.

Formally, given a set of users $U = \{u_1, \ldots, u_m\}$, and a set of items $I = \{i_1, \ldots, i_m\}$, we use a set $V$ to represent the main entities of a ranking-oriented recommender systems and define it as $V = <U, P, R>$ where $U$ is the set of users, $R = \{(i_d, i_u) | i \in I\}$ is the set of representatives denoting desirable and undesirable sides of items, and $P = \{<i, j>, i \in I, j \in I\}$ is the set of pairwise preferences. Note that each $p = <i, j>$ indicates preference of $i$ over $j$. For legibility, we call the first item as $p$'s desirable item, denoted by $p^d$, and the second one as $p$'s undesirable item denoted by $p^u$. The agreement function $agg$ and support function $sup$ define the relation between different entities of ranking-oriented recommender systems.

**Definition 1.** Given a user $u_i \in U$, and a preference $p_j \in P$, the agreement function $agg: U \times P \to \{0,1\}$ indicates the relations between users and pairwise preferences:

$$agg(u_i, p_j) = \begin{cases} 1, & \text{if } <u_i, p_j^d, p_j^u> \in O \\ 0, & \text{otherwise} \end{cases}$$

where $O$ is the observation set of preferences.

**Definition 2.** Given a preference $p = (i, j)$ and a representative $r$, the support function $sup: P \times R \to \{0,1\}$ indicates whether a preference $p$ supports the representative $r$ or not.

$$sup(p, r) = \begin{cases} 1, & \text{if } p^u = i \text{ and } r = i_u \\ 1, & \text{if } p^d = i \text{ and } r = i_d \\ 0, & \text{otherwise} \end{cases}$$

Using agreement and support functions, we can construct the tripartite preference graph (TPG) of a pairwise preference dataset as in Definition 3 [20]. Fig.1 shows the network schema of TPG. Users are connected to preferences which they agree, and, representatives are connected to those preferences that supports them.

**Definition 3.** A **tripartite preference graph** is a tripartite graph $TPG = <(U, P, R), (E_{UP}, E_{PR})>$ where U, is the set of users, P is the set of possible pairwise preferences, and R is the set of items' representatives. $E_{UP} = \{(u, p) | agg(u, p) = 1, u \in U, p \in P\}$, is the set of edges that connect a user to a preference if there is an agreement relation between that user and preference. Also, $E_{PR} = \{(p, r) | sup(p, r) = 1, p \in P, r \in R\}$, is the set of edges linking each pairwise preference to the representatives that it supports.

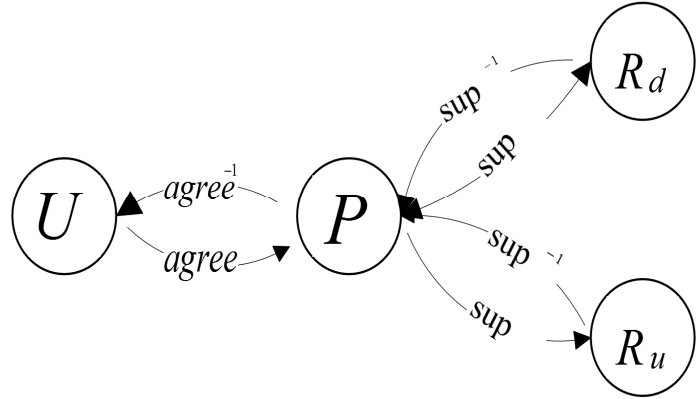

Fig.1. Network schema of tripartite preference graph (TPG)

More clearly, TPG is a tripartite graph which models the subjective relations between users and items' representatives through intermediate nodes of pairwise preferences. This intermediate layer provides an ability to model the choice context in which a user has preferred/not preferred an item (See Fig.2c). Moreover, the items' representatives layer enables TPG to model different sides of an item $i$: when $i_d$ reflects the situation in which $i$ is preferred over another item, and $i_u$ refers to the situation in which another item is preferred over $i$. Fig.2 illustrates a schematic example of TPG and its difference with tradition bipartite graph (BG) representation of recommender systems. As illustrated in Fig.2b, BG only represents the interaction between users and items, and so, Martin and Lee both are connected to $B$ in a similar way while they have picked item $B$ in different contexts. On the other hand, TPG can clearly reflect that these users do not agree over any pairwise comparison.

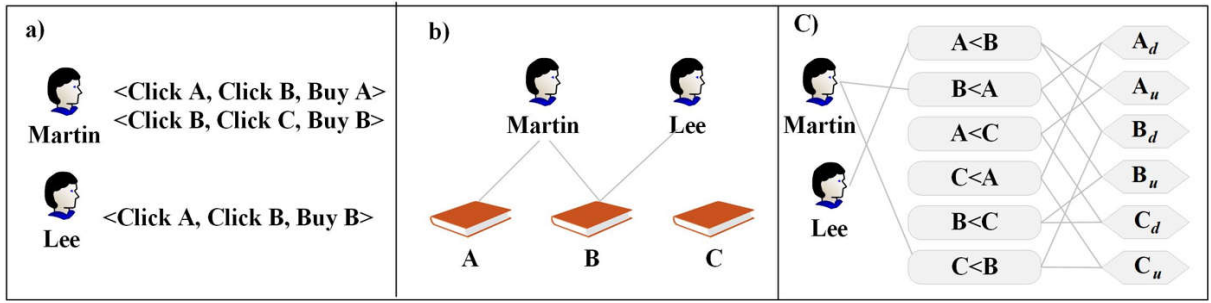

Fig.2. BG and TPG representation of a pairwise preference dataset.

Given the graph representation of a recommender system, graph-based recommendation algorithm should define a function *score*: $t_q \times t_t \to \mathbb{R}$ in order to reply queries in form of query $Q = (t_q, t_t)$ where $t_q$ is the object type of query nodes and $t_t$ is the object type of target nodes. More clearly, the *score* function calculates scores for objects with type $t_t$ based on the degree of closeness of those objects to some particular objects with type $t_q$ through paths that we call recommendation paths, as in Definition 4.

**Definition 4.** Given a network $G$ and a query format $Q = (t_q, t_t)$, any path connecting a node $x \in t_q$ to another node $y \in t_t$, *is called a recommendation path over G with respect to Q.*

Graph-based collaborative ranking algorithms seek to reply the query in forms of $Q = (U, R)$ and score representatives according to their closeness to the target user. Therefore, ranking – oriented recommendation paths are those paths that connect a user node to a desirable or undesirable representative in graph representation of ranking-oriented recommender systems. As this paper focuses on graph-based collaborative ranking frameworks, we simply refer ranking-oriented recommendation paths as recommendation paths over $G$.

As stated earlier, a main challenge that a graph-based recommendation algorithm faces is to determine which recommendation paths should be taken into account by the scoring function. In the following sections, we will follow a systematic approach to determine reliable recommendation paths for different classes of neighbor based collaborative ranking and we will show how to use these reliable recommendation paths to score items more precisely.

## 4. String description of recommendation meta-paths

In this paper we seek to determine which recommendation paths in TPG are reliable. Clearly, it is not feasible or useful to individually analyze the reliability of each recommendation path between any pair of user and representative nodes. So, we use a high-level description of recommendation paths, called recommendation meta-paths.

Let $G = <V, E>$ be a heterogeneous network and $SG = (V_{SG}, E_{SG})$ be its network schema [27] in which $V_{SG} = \{type(v) | v \in V\}$ contains the object types of G's vertices, and $E_{SG} = \{(type(i), type(j)) | (i,j) \in E\}$. The set of network meta-paths are defined as the whole set of paths in the network and are denoted by *as* $\Psi = \{\psi = (v_1, .., v_n) | \forall_{1 \leq i < n} (v_i, v_{i+1}) \in E_{SG}\}$ on the graph of network schema. We note that each meta-path $\psi$ provides a high level description of one possible type of paths over $G$ as a heterogeneous network. For instance, the UPR meta-path is a path over TPG's schema that starts at $U$, passes its outgoing edge to $P$ and then passes an edge to reach R. This meta-path describes all paths over TPG that connect a user to a representative node through a preference node. The *UPR* meta-path describes paths such as $< Lee - A < B - A_u >$ and $< Martin - C < B - B_d >$ as illustrated in Fig2.c.

The concept of meta-paths in heterogeneous networks enables us to describe the semantics behind different relations among entities of recommender systems [8,9,21,22,27,35]. Motivated by the concept of meta-path, we define the concept of recommendation meta-paths with regards to the query format $Q = (t_q, t_t)$ over the network $G$ in Definition 5.

**Definition 5.** A *recommendation meta-path* $\phi$ in a graph $G$ with respect to a query $Q = (t_q, t_t)$ is a path defined on $G$'s network schema $SG = (V_{SG}, E_{SG})$ from $t_q$ to $t_t$ where $t_q$ and $t_t$ are the node types corresponding to the query and the target objects of recommendation. Also, $\Gamma = \{\phi = $

$(v_1, \ldots, v_n) | v_1 = t_q, v_n = t_t, \forall_{1 \leq i < n} (v_i, v_{i+1}) \in E_{SG}\}$ is the set of all recommendation meta–paths existing in $G$.

Different recommendation meta-paths can model different approaches for recommendation. For instance, in a heterogeneous networks comprised of users, movies, actors and relations between them, the $< User - User - Movie >$ meta-path can be used to recommend a movie to a user because his friend has watched it. This approach is exactly what is used in user-based recommendation. On the other hand, the $< User - Actor - Movie >$ meta-path can help us to relate a target user to a movie in which his favorite actor has played a role. This latter approach is an example of content-based class of recommendation methods.

Typically, there are a large number of recommendation meta-paths that exist in a graph and some of them are reliable when using a certain NC-Rank approach. Here we define *the string description* of the recommendation meta-path set $\Gamma$, denoted by $strdesc_\Gamma$, to shortly demonstrate the patterns appearing in meta-paths that belong to $\Gamma$.

The string description of meta-paths in a graph $G$ are obtained from a finite alphabet $\Sigma = \{t_1 \ldots t_o\}$ where $\{t_1, \ldots, t_o\}$ represents the set of node types in $G$. Each member $t_i$ of $\Sigma$ describes a meta-path $t_i$ in G that is a meta-path with length zero from a node type $t_i$ to $t_i$ that is conceptually equivalent to starting from node with type $t_i$ and staying there and not moving anywhere. Also a meta-path $\phi = (v_1, v_2, \ldots, v_n)$ is described by a string $s = v_1 v_2 \ldots v_n$ that represents starting from a node with type $v_1$, following the meta-path $\phi$ that end in a node with type $v_n$. For instance, the meta-path $\phi = (U, P, R, P, U)$ over TPG is described by string $UPUPU$ while string U describes the meta-path $\phi = (U)$.

We define a set of operations that are needed to systematically obtain complicated string descriptions from simple ones. Given two string descriptions of meta-paths $\alpha = \{t_1 \ldots t_m\}$ and $\beta = \{t'_1 \ldots t'_n\}$.

- **Select operation** denoted "|", is used to describe different meta-path options between nodes. For example if $t_1 = t'_1$ then $e = (\alpha | \beta)$, is a string description that represents that one of meta-paths α or β may be selected to be followed from a node with type $t_1 (= t'_1)$. For example, if $\alpha = UPU$ and $\beta = UPRPU$, α|β would be represented by the string description $s = UPU|UPRPU$ and describes one of UPU or UPRPU paths can be selected for connecting two user nodes (that are nodes with type U).

- **Join Operation**, denoted by ".", is used to describe concatenating two meta-paths. For example if $t_m = t'_1$, then α.β is string description for a meta-path that connects nodes of type $t_1$ to nodes of type $t'_n$ through following a meta-path that is described by β just after finishing a meta-path described by α. Formally, we have $\alpha.\beta = [t_1 \ldots t_m].[t'_2 \ldots t'_n] = (t_1 \ldots t_m t'_2 \ldots t'_n)$. For instance, if we have $\alpha = UP$ and $\beta = PR$, then α.β would be equal to $UPR$.

- **Repeat Operation**, denoted by " ∗ ", is used to describe meta-paths that are generated by repeatedly following some meta-path α, for zero or more times. It is applicable to the string description of a meta-path only if the starting and ending node types of that meta-path are the same. For example if $t_1 = t_m$, then $\alpha^*$ describes a meta-path that starts from and ends at a certain object type $t_1$, by following meta-path described by α, zero or more times. Formally, $\alpha^* = \{\alpha^0, \alpha^1, \dots, \alpha^n, \dots\}$, and since α is a string description for a meta-path starting at a node type $t_1$, then $\alpha^0$, that means following α for zero times, denotes not moving from $t_1$ to anywhere. Also string $\alpha^n$ is generated by applying *join operation* over α for $n-1$ times. In other words, $\alpha^*$ refers to the strings that belongs to the set $S = \{t_1, t_1 \dots t_m, t_1 \dots t_m t_2 \dots t_m, \dots\}$. For example, if $\gamma = UPU$ then $\gamma^* = [UPU]^* = \{U, UPU, UPUPU, ..\}$.

To make the whole concept clear, let us define the string description of recommendation meta-paths of GRank denoted by $strdesc_{GRank}$. GRank seeks to answer the query $Q = (U, R)$ as it seeks to determine the closeness of representatives to a given target user. For this purpose, it applies personalized PageRank over TPG, and so it considers all recommendation meta-paths in TPG. The set of TPG's recommendation meta-paths can be obtained through Breadth-first search on network schema starting at user node $U$. These meta-paths first pass edges that are described by $\alpha_1 = UP$ to reach a preference node. Then, it leaves P and returns to it passing edges in the form of $\alpha_2 = PU$ and $\alpha_1 = UP$ or $\alpha_3 = PR$ and $\alpha_4 = RP$. These paths are described by string in form of $[\alpha_2.\alpha_1|\alpha_3.\alpha_4] = [PUP|PRP]$. This pattern can be repeated for arbitrary number of times which is expressed by $[PUP|PRP]^*$. Finally, these paths will end through passing the meta-path $\alpha_3 = PR$ to reach a representative node R. Therefore, $strdesc_{GRank} = \alpha_1.[PUP|PRP]^*.\alpha_3 = UP.[PUP|PRP]^*.PR$.

**Lemma. 1.** *Given two string descriptions $\alpha = \{t_1 \dots t_m\}$ and $\beta = \{t'_1 \dots t'_n\}$ where $t_1 = t_m = t'_1$, we have $[\alpha^*].[\beta] = (t_1 \dots t_m)^* t'_2 \dots t'_n$*

**Proof:** *From the definition of the repeat operation, we know that $[\alpha^*]$ ends either with a $t_1$ or a $t_m$ that are the same and both are equal to $t'_1$. So, $[\alpha^*] = (t_1 \dots t_m)^* = [\dots t_m]$ and $[\beta] = t_m t'_2 \dots t'_n$, and according to the definition of the join operation, we have $[\alpha^*].[\beta] = (t_1 \dots t_m)^* t'_2 \dots t'_n$.*

□

**Lemma.2.** *Given two string dsescriptions of meta-paths $\alpha = \{t_1 \dots t_m\}$ and $= \{t'_1 \dots t'_n\}$ where $t_1 = t_m = t'_1$, we have $.[\beta].[\alpha^*] = t'_1 \dots t'_{n-1}(t_1 \dots t_m)^*$*

**Proof:** The proof is similar to that of Lemma.1.

□

As we showed that $strdesc_{GRank} = UP.[PUP|PRP]^*.PR$. Using Lemma.1 and Lemma.2 we can further simplify the string description to see that $strdesc_{GRank} = U[PUP|PRP]^*R$.

Using string description of meta-paths, we can shortly describe the reliable recommendation meta-paths of NC-Rank algorithms and also check whether the recommendation paths of an algorithm, like GRank, are consistent to string description of reliable recommendation meta-paths of a particular class of NC-Rank algorithms or not.

## 5. Discovering reliable meta-paths for neighborhood collaborative ranking

Graph-based collaborative ranking algorithms should define a ranking function $rank: U \times I \to \mathbb{R}$, which scores each item based on the degree of closeness of its corresponding representatives to the target user. More formally, a graph-based collaborative ranking algorithm, $alg$, needs to define a ranking score function as:

$$rank(u, i) = f(score_{alg}(u, i_d), score_{alg}(u, i_u)) \qquad (1)$$

Where $i_d$ and $i_u$ are desirable and undesirable representatives for an item i, and $score_{\Gamma_{alg}}(u, r)$ is a function that measures the degree of closeness between a user $u$ and a representative $r$ based on recommendation meta-paths that are considered reliable in the recommendation algorithm $alg$. In this setting, function f must be defined to aggregate the scores of representatives of each item $i$ in order to estimate the overall desirability of i for user u.

The first step for defining such a scoring function is to find the reliable recommendation meta-paths for an algorithm $alg$. In this section we are going to first define the concept of similarity or closeness that is used in a neighborhood collaborative ranking method $alg$. We will also show how to describe all recommendation meta-paths that are in harmony with that concept of similarity, using a string description $strdesc_{alg}$. Here, we will do that for user-based, preference-based and representative based approaches to the neighborhood collaborative ranking. Finally, we show that GRank uses some recommendation meta-paths that are not in harmony with the rules of any of the three mentioned approaches, and can be considered as unreliable meta-paths.

### 5.1. User-based neighborhood collaborative ranking

Like neighborhood collaborative rating algorithms, user-based neighborhood collaborative ranking (UNC-Rank) algorithms estimate $score(u, r)$ using opinions of $u$'s neighbors about $r$. More clearly, UNC-Rank algorithms define the scoring function according to the following rules:

- ***Rule U-1:*** A representative is close to the target user if it is voted directly by him or his similar users. A user $u$ votes to a representative $r$ if he agrees with some pairwise preference that supports r.
- **Rule U-2:** A user is similar to the target user if he is directly similar to him or is similar to users that are similar to the target user.

Now, the question is how to calculate direct users' similarities using TPG. TPG reflects the similarity over pairwise preferences as well as the similarity over representatives[20]. The first one that we call type-1 user-user similarity, indicates that two users are similar if they agrees over some pairwise preferences. For instance, Jack and John are similar as they both have preferred B over A, and, D over A (See Fig.3a). This similarity can be determined through $<John, A < B, Jack>$ and through $<John, A < D, Jack>$. Trivially, this type of paths are described by UPU meta-paths. On the other hand, type-2 user-user similarity demonstrates that two users are similar if they agree upon the overall desirability/undesirability of items for them. For instance, in Fig. 3b, we can infer that Jack is similar to John more than Lee as Jack and John both have preferred some item over A while Lee has preferred $A$ over $D$. We can infer similarity between Jack and John through the path of $<Jack, A < B, A\_u, A < C, John>$ that is described by meta-path UPRPU over TPG. Type-1 and Type-2 direct user-user similarities can be respectively obtained through meta-paths $UPU$ and $UPRPU$ in TPG's schema. We repeat that UPU is a meta-path over TPG as it is a valid path over TPG's schema; it starts at U, then passes an outgoing edge to reach P, and finally it returns to U.

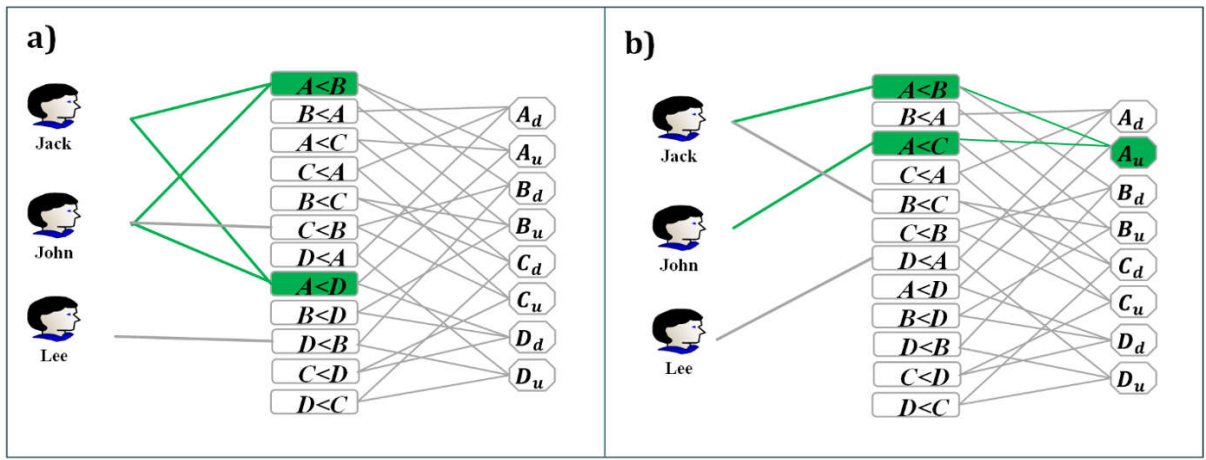

Fig.3. A schematic example to illustrate different types of direct similarities among users. a) based on pairwise preferences. b) based on desirability/undesirability of items.

According to recursive definition of users' similarities (Rule. U-2), and scoring logic (Rule. U-1), we can define the string description of user-based reliable recommendation meta-paths as strdesc$_{UNC}$=$(UPU|UPRPU)^*PR$ (See Definition 6). Note that $UPU$ and $|UPRPU$ denotes two alternatives to find users that are directly similar to the target user, and, its repetition $(UPU|UPRPU)^*$ describes meta-paths between users that are directly /indirectly similar to each other. We also mention that meta-path UPR connects users to representatives that are directly voted by them. Therefore, the concatenation of $(UPU|UPRPU)^*$ and $R$, denoted by $(UPU|UPRPU)^*PR$, describes meta-paths that are directly voted by the target user or users that are directly/indirectly similar to him. Therefore, UNC-Rank algorithms should estimate $score\ (u,r)$ through paths that are consistent with $strdesc_{UNC}$ as in Definition 6.

**Definition 6.** $strdesc_{UNC}$ is the string description describing the reliable recommendation meta-paths for user-based NC-Rank algorithms. $strdesc_{UNC} = (UPU|UPRPU)^*PR$ indicating paths that start at the target user $u$, then meet similar users to the target user based on type-1 or type-2 similarities) i.e. following meta-paths in form of $(UPU|UPRPU)^*$ and then passing the meta-path UPR

## 5.2. Preference-based neighborhood collaborative ranking

Preference-based neighborhood collaborative ranking takes into account similarities among pairwise preferences to define the scoring function. Preference-based algorithms define the scoring function using these rules:

- **Rule P-1.** A representative is close to a user if it is supported by his concordant preferences.
- **Rule P-2.** A preference $p$ is concordant to the target user $u$ if $u$ agrees with $p$ or agrees with preferences similar to $p$.
- **Rule P-3.** A preference $p'$ is similar to preference $p$ if it is directly similar to $p$ or is similar to preferences that are similar to $p$.

The next step is to define direct similarity among pairwise preferences. Like items' similarities in collaborative rating algorithms, we assume that two pairwise preferences, $p_i$ and $p_j$ are directly similar if there are (ideally a large number of) users who agree with both $p_i$ and $p_j$. Fig. 4a depicts a schematic example of two pairwise preferences $(B < A)$ and $(B < D)$ that are similar, as users that prefer $A$ over $B$, also prefers $D$ over $B$. On the other hand, the pairwise preferences $(C < A)$ and $(A < D)$ are dissimilar as there is no paths connecting them (See Fig.4b) since no user agrees with both of these preferences. These similarities can be captured through PUP paths in TPG, that can be extended to find indirect similar preference through repetitions of PUP paths, expressed by $[PUP]^*$

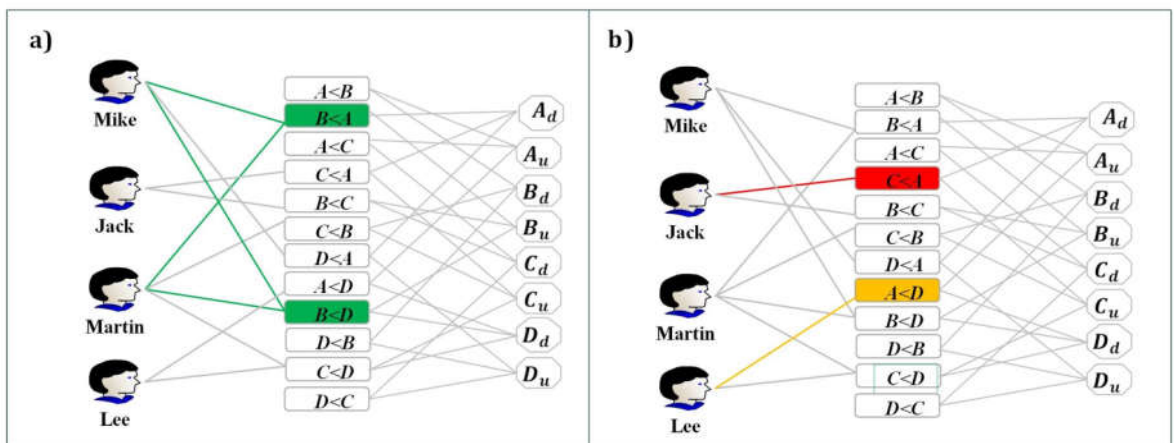

Fig.4. A schematic example to illustrate similarity/dissimilarity between pairwise preferences.

According to Rule P.1-3, the PNC-Rank algorithms make recommendations according to paths that are described through concatenation of $\alpha = UP$, $\beta = [PUP]^*$, and $\gamma = PR$. Therefore, we can define $strdesc_{PNC} = U[PUP]^*R$ as mentioned in Definition 7.

**Definition 7.** $strdesc_{PNC}$ refers to the string description of the reliable recommendation meta-paths of preference-based NC-Rank algorithms. We define $strdesc_{PNC} = U[PUP]^*R$ that is generated through applying join operations over meta-path $UP$, $[PUP]^*$, and PR. Note that UP meta-paths reach users to their direct concordant preferences, $[PUP]^*$ enable us to find preferences that directly/indirectly similar to his concordant preferences, and finally, PR determines representatives that are supported by these direct/indirect concordant preferences.

**5.3. Representative-based neighborhood collaborative ranking**

Representative-based neighbor-based collaborative ranking defines the scoring method using these rules:

- **Rule R-1:** A representative is close to a target user $u$ if it is similar to representatives that $u$ has voted to them
- **Rule R-2:** A representative $r'$ is similar to representative $r$ if it is directly similar to $r$ or is similar to representatives that are similar to $r$.

Similar to preference similarity, we assume that two representatives are directly similar if they have been voted by similar, ideally equal, sets of users. For instance, in Fig. 5a, the representatives $A_d$, and $D_d$ are similar as several users have voted both of them, and ranked them above $B$ and $C$. Nevertheless, $B_d$, and $C_d$ seem dissimilar in Fig. 5b, as there is no user voting for both of them. These similarities can be captured through repetition of meta-path $(RPURP)$ in TPG, and so, the recommendation paths of representative-based algorithms are described by $strdesc_{RNC} = UP(RPUPR)^*$ as stated in Definition 8:

**Definition 8.** $strdesc_{RNC}$ denotes the string description denoting the reliable recommendation meta-paths from representative-based NC-Rank perspective. We define $strdesc_{RNC} = UP(RPUPR)^*$ that is generated through joint of UPR and $(RPUPR)^*$ meta-paths that first connect users to the representatives that are directly voted by the target user, and then, pass meta-paths in form of $(RPUPR)^*$ to find directly/indirectly similar representatives.

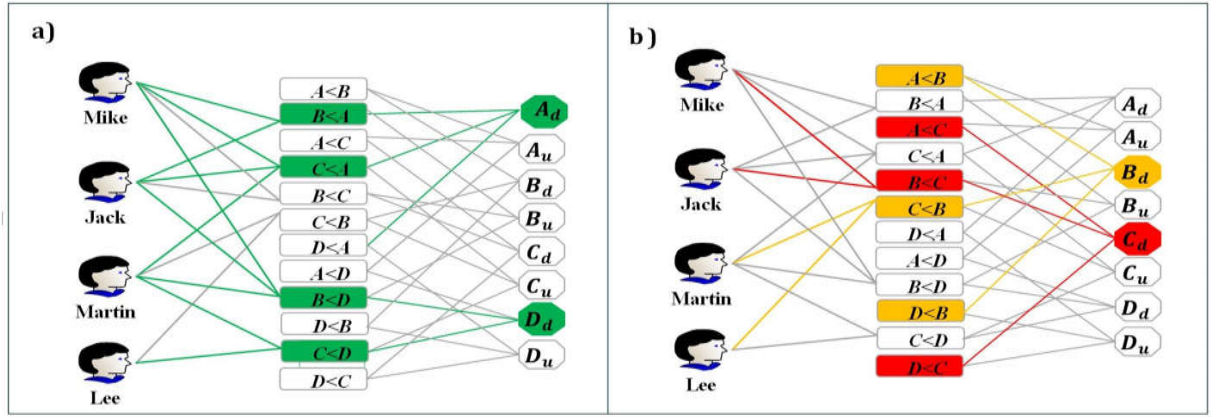

Fig. 5. A schematic example to illustrate similarity/dissimilarity among representatives.

### 5.4. Reliability analysis of GRank's recommendation meta-paths

Once the string descriptions of NC-Rank methods are extracted, we can answer the question that if GRank scores items through reliable recommendation meta-paths or not. As mentioned before, GRank applies personalized PageRank to score items through paths described by, $strdesc_{GRank} = U[PUP|PRP]^*R$ , that refers to all existing paths in TPG. For instance, in Fig.6, GRank might score items for Jack through the path $\pi = \{Jack, A < B, B_d, C < B, C_u, C < A, A_d\}$ that follows meta-path $\rho = \{UPRPRPR\}$ which is not compatible with any of the reliable recommendation meta-paths described by $strdesc_{UNC}, strdesc_{PNC}$ or $strdesc_{RNC}$. To see why, please notice that sequence RPR appears in $\rho$ while it cannot appear in any of these reliable recommendation meta-paths. Paths such as $\pi$, that pass a RPR meta-path, implicitly assume that two representatives are similar if they are connected to a common pairwise preference. Trivially, that does not seem reasonable as it implies that all desirable representatives are similar to all undesirable representatives. Note that each item has been compared to all other items in the preference layer, and so, each desirable item's representative $i_d$, is connected to the undesirable representatives of other items through RPR meta-paths. Therefore, inferring similarity through these paths will mislead the ranking algorithm. For instance, in the mentioned example, $\pi$ connects Jack to $A_d$ because it implicitly assumes that $B_d$ is similar to $C_u$ and , $C_u$ is similar to $A_d$. It is clear that voting for $A_d$ is trivially in contrast with the original opinion of Jack who prefers C over A. Trivially, considering such paths that are not compatible with the desired meaning of similarity or closeness in neighborhood recommendation, may deviate the ranking algorithm from the real interests of the users and avoid it from making accurate recommendations to them. Therefore, GRank that applies personalized PageRank over TPG and considers all existing meta-paths, is not totally consistent with any of the basic approaches in neighborhood collaborative ranking.

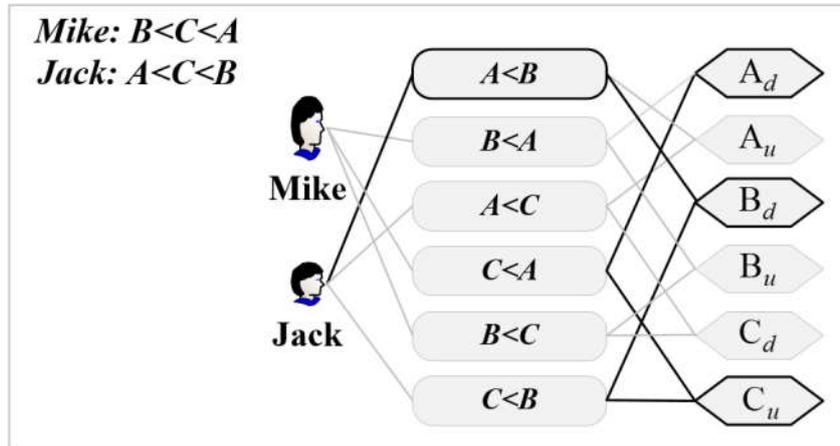

Fig.6. A schematic example to illustrate invalid meta-paths in TPG.

## 6. Top-N recommendation through reliable meta-paths for neighborhood collaborative ranking

So far, we have defined the string description, $strdesc_{alg}$ that describe $\Gamma_{alg}$, the set of recommendation meta-paths that are reliable from the viewpoint of user-based, preference-based, and representative-based recommendation. Here, we propose a novel collaborative ranking framework, called ReGRank, that scores items through reliable recommendation meta-paths of TPG.

Personalized PageRank[15] is one of the most acknowledged proximity metrics which weighs the whole set of paths in a network to assign a proximity value to a pair of nodes consisting a target node and some other node in the network[8]. Personalized PageRank models the behavior of a random surfer who starts at a node corresponding to the target user, and then randomly surfs the network to reach other entities. The random surfer will increase the rank of each node every time that he reaches that node. Trivially, the higher ranks are assigned to the nodes that are more reachable from the target user. That make sense in recommender systems as they seek to find items that there are many evidences for relating them to the target user [8,9,19]. However since there are some unreliable meta-paths in the original TPG network, it is not safe to use Personalized PageRank directly on TPG. On the other hand, it is hard to individually consider each reliable recommendation meta-path($\rho \in \Gamma_{alg}$) and find each instance of that meta-path in TPG for calculating the overall proximities among nodes, because $\Gamma_{alg}$ is an infinite set containing a large number of recommendation meta-paths. Therefore, ReGRank projects TPG to graph $g_{alg} =< (U,R), E >$ which contains all reliable meta-paths described by $strdesc_{alg}$. Thereafter, it applies personalized PageRank on $g_{alg}$ to rank items using the reliable meta-paths that have been defined for that class of recommendation (e.g. user-based).

### 6.1. Projecting TPG through reliable recommendation meta-paths

Generally, network projection is used to compress a network of *n* types of nodes to another network with *l* different types of nodes where $l < n$. That provides an ability to directly show the relations between a particular set of objects with possibly different types. Here, we are going to project TPG to a 2-types graph consisting users and representatives as we are interested in recommendation meta-paths that start at users and ends at representatives. To project TPG, we should determine which types of relations should be kept in the projected graph and also, how to weigh each relation in the projected graph.

### 6.1.1. Projection strategy

As we mentioned there are different approaches to neighborhood recommendation, each defining a different set of meta-paths, $strdesc_{UNC}$, $strdesc_{PNC}$ and $strdesc_{RNC}$, and here we are going to construct projected graphs each containing meta-paths that reflect one of those approaches to recommendation. In other words, each projected networks should contain all paths described by one of these string descriptions and no other recommendation meta-paths. For this purpose, we first define how to project a network through a set of meta-paths in Definition 9. For instance, if $\theta_{alg} = \{UPR, UPU\}$, the projected network of Fig. 7a will be like the one shown in Fig. 7b; Martin and Mike are connected to each other as they have connected through paths $\{< Mike, B < A, Martin >, < Mike, B < D, Martin >\}$ that represent paths between Mike and Martin that are consistent with meta-path $UPU$. Note that Jack is not connected to other users due to absence of any UPU paths between Jack and them. Moreover, each user is connected to representatives that he has voted for. As an example, Mike is connected to $A_d$ through path $< Jack, C < A, A_d >$ that follows meta-path $UPR$. Note that there is no edge from representatives to users as we have projected TPG over the meta-path set $\theta_{alg} = \{UPR, UPU\}$ in which there is no meta-paths from representatives to users. Furthermore, the projected network contains two directed links, associated with a weight, between users who have a *UPU* path in TPG.

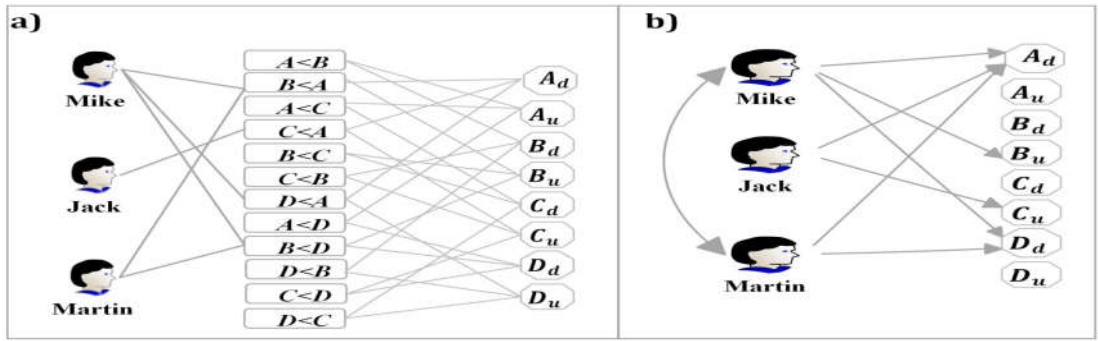

Fig.7. A schematic example to explain the compressing mechanism through a set of meta-paths in $\theta_{alg} = \{UPR, UPU\}$. a) the original network b) the projected network

**Definition 9.** Projection of TPG through a set of meta-paths $\theta_{alg} = \{\pi_1, \dots, \pi_m\}$ generates a graph with network schema $net =< V, E >$ where $V = \{\cup_{i=1}^{i=n} \pi_i.start\} \cup \{\cup_{i=1}^{i=n} \pi_i.end\}$ and $E =$

$\{(\pi_i.start, \pi_i.end)|\pi_i \in \theta_{alg}\}$ where $\pi_i.start$ and $\pi_i.end$ indicate the node types for the starting and ending nodes of $\pi_i$, respectively.

Prepositions 1-3 state that projection of TPG with regards to a set of meta-paths $\theta_{UNC}$, $\theta_{PNC}$ and $\theta_{RNC}$ will generate three networks, and the set of recommendation meta-paths for each of those networks is described by $strdesc_{UNC}$, $strdesc_{PNC}$, and $strdesc_{RNC}$, respectively. Therefore, applying personalized PageRank of the target user over these networks will score items according to meta-paths that are described by $strdesc_{UNC}$, $strdesc_{PNC}$, or $strdesc_{RNC}$, respectively.

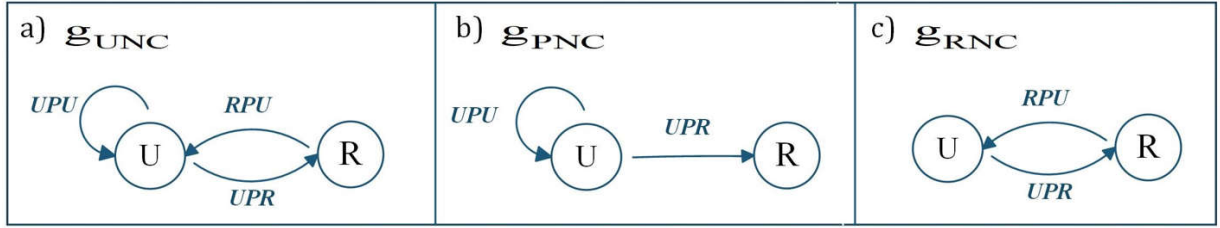

Fig. 8. Network schema of TPG's transformation for UNC-Rank, PNC-Rank, and RNC-Rank

**Preposition.1.** *Projecting TPG over meta-paths in $\theta_{UNC} = \{UPU, UPR, RPU\}$, results in a graph $g_{UNC}$, and the set of all all recommendation meta-paths in $g_{UNC}$, from users to representatives, in $g_{UNC}$, is equal to the set of all reliable meta-paths described by $strdesc_{UNC} = (UPU|UPRPU)^*PR$*

**Justification:** If we project TPG over $\theta_{UNC} = \{UPU, UPR, RPU\}$, we obtain a graph with network schema of Fig. 8a. According to Definition 9, the projected network will contain two types of nodes, that are user and representative nodes, and three types of edges, between users that are connected through UPU paths, from users to representatives based on UPR paths, and from representatives to users based on RPU paths. Therefore, the network schema will connect edges from U to R, R to U, and U to U as in Fig. 8a.

These set of all recommendation meta-paths of a network can be obtained through breadth first search over its schema starting at U node[27]. The paths start at node U, then return to U passing meta-path (UU) or through passing meta-path *UR* and RU. These patterns show two alternatives for moving from U to U, and so, can be expressed by $(UU|URU)$. This loop can be repeated for a number of times, and so we use $(UU|URU)^*$ expression to describe paths from user nodes to each other in $g_{UNC}$. Additionally, these paths can be joined with the UR meta-path to form a recommendation meta-path from users to representatives. Therefore, these recommendation meta-paths can be described by $strdesc_{g-UNC} = [(UU|URU)^*].[UR] = (UU|URU)^*R$.

The remaining step is to prove that the recommendation meta-path set over $g_{UNC}$, described by $strdesc_{g-UNC}$, is equivalent to the set of reliable recommendation meta-path set over TPG that is described by $strdesc_{UNC}$. More formally, we should prove that $\{(UU|URU)^*R\}_{g-UNC} \cong \{(UPU|UPRPU)^*PR\}_{TPG}$

Aforementioned, UR and RU edges in $g_{UNC}$ respectively refer to UPR and RPU paths in TPG. So, we can infer that $\{URU\}_{g-UNC} \cong \{UPRPU\}_{TPG}$. Additionally, we know that UU edges in $g_{UNC}$ refer to UPU paths in TPG. Therefore, we can infer that $\{(UU|URU)^*\}_{g-UNC} = \{(UPU|UPRPU)^*\}_{TPG}$. Furthermore, as $(UU|URU)^*R$ is the obtained through concatenation of $\{(UU|URU)^*\}_{g-UNC}$ with UR, we should join $\{(UPU|UPRPU)^*\}_{TPG}$ to $\{UPR\}_{TPG}$ to form equivalent of $\{(UU|RU)^*R\}_{g-UNC}$ over TPG. So, we can infer that

$$strdesc_{g-UNC} = \{(UU|URU)^*R\}_{g-UNC}$$
$$\cong [((UU|URU)^*].[UR]_{g-UNC} = [(UPU|UPRPU)^*].[UPR]_{TPG}$$
$$= \{(UPU|UPRPU)^*PR\}_{TPG} = strdesc_{UNC}$$

□

**Corollary. 1.** *Projection TPG with regards to the set of meta-paths $\theta_{UNC}$ will result in network $g_{UNC}$. Personalized PageRank of the target user in $g_{UNC}$ only takes into account the reliable meta-paths in user-based approach to NC-Rank.*

**Preposition. 2.** *Projecting TPG over meta-paths in $\theta_{PNC} = \{UPU, UPR\}$, result in a graph $g_{UNC}$ whose all recommendation meta-paths that are described by $strdesc_{PNC} = U(PUP)^*R$,*

**Justification:** Projecting TPG over $\theta_{PNC} = \{UPU, UPR\}$ will generate a network in which users are linked to each other if there is a UPU path between them in TPG, and also, a user will be linked to a representative in case that there exist a UPR path between them in TPG. Accordingly, the projected graph will contain user-user and user-representative edges and its schema would be as in Fig. 8b.

Similar to preposition.1, we can see that the recommendation paths of $g_{PNC}$ are described by $strdesc_{g-PNC} = \{(UU)^*R\}$ because the recommendation meta-paths contain repetitions of loops from U to U for some number of times and then they all pass the UR meta-path to form a recommendation meta-path from users to representatives.

As $\{UR\}_{g-PNC} \cong \{UPR\}_{TPG}$, and, $\{UU\}_{g-PNC} \cong \{UPU\}_{TPG}$, we can infer that $\{(UU)^*\}_{g-PNC} \cong \{(UPU)^*\}_{TPG}$.

Consequently, we have:

$$\{(UU)^*R\}_{g-PNC} = [(UU)^*].[UR]\}_{g-PNC} \cong \{[(UPU)^*].[UPR]\}_{TPG} = \{(UPU)^*PR\}_{TPG}.$$

Additionally, we have

$$\{(UPU)^*PR\}_{TPG} = \{UPR, UPUPR, UPUPUPR, \ldots\}$$
$$= \{U(PUP)^0R, U(PUP)^1R, U(PUP)^2R, \ldots\}$$
$$= \{U(PUP)^*R\}_{TPG}$$

So, we can infer that

$$strdesc_{g-PNC} = \{(UU)^*R\}_{g-PNC} \cong \{(UPU)^*PR\}_{TPG} = \{U(PUP)^*R\}_{TPG} = strdesc_{PNC}.$$

☐

**Corollary 2.** *Projection TPG with regards to the set of meta-paths $\theta_{PNC}$ will result in network $g_{PNC}$. Personalized PageRank of the target user in $g_{PNC}$ only takes into account the reliable meta-paths in preference-based approach to NC-Rank.*

**Preposition 3.** Projecting TPG over meta-paths in $\theta_{RNC} = \{UPR, RPU\}$, results in a graph $g_{UNC}$ whose *all recommendation meta-paths are described by $strdesc_{RNC} = UP(RPUPR)^*$,*

**Justification**: Projecting TPG over $\theta_{RNC} = \{UPR, RPU\}$ results in a network in which users and representatives are connected to each other if there is a path in form of UPR/RPU between them. The projected network, denoted by $g_{RNC}$, is a bipartite directed network with a schema of Fig. 8c.

Aforementioned, all recommendation meta-paths of $g_{RNC}$ are obtained through applying breadth first search over its schema. Trivially, these paths contain the UR link and some repetition of loops from R to R that is described by the string in form of $(RUR)^*$. Therefore, the recommendation meta-paths of $g_{RNC}$ denoted by $strdesc_{g-RNC} = U(RUR)^*$ that is generated by applying a join operation over UR and $(RUR)^*$,

As $\{UR\}_{g-RNC} \cong \{UPR\}_{TPG}$, and $\{RU\}_{g-RNC} \cong \{RPU\}_{TPG}$, we can infer that $\{RUR\}_{g-RNC} \cong \{RPUPR\}_{TPG}$. Furthermore, as $\{U(RUR)^*\}_{g-RNC}$ is generated through concatenation of UR with $(RUR)^*$, its equivalent over TPG will be obtained joining $\{UR\}_{TPG}$ and $\{(RPUPR)^*\}_{TPG}$. So, we have

$$strdesc_{g-RNC} = \{U(RUR)^*\}_{g-RNC}$$
$$= \{[UR].[(RUR)^*]\}_{g-RNC} \cong \{[UPR].[(RPRPU)^*]\}_{TPG}$$
$$= \{UP(RPUPR)^*\}_{TPG} = strdesc_{RNC}.$$

☐

***Corollary. 3.*** *Projection TPG with regards to the set of meta-paths $\theta_{RNC}$ will result in network $g_{RNC}$. Personalized PageRank of the target user in $g_{RNC}$ only takes into account the reliable meta-paths in representative-based approach to NC-Rank.*

It is worth mentioning that there is an interesting relation. between different network schemas; $g_{PNC}$ takes into account relations over pairwise preferences, but it does not provide an ability to propagate ranking from representatives. On the other hand, $g_{RNC}$ models users' similarity over representatives but it ignores the relations between users and preferences. $g_{UNC}$ is where $g_{RNC}$ and $g_{PNC}$ meets each other: it lets similarities to be calculated based on both available evidences, common pairwise preferences and common representatives among users.

### *6.1.2. Weight assignment*

Now that we projected TPG using a set of meta-paths $\theta_{alg}$ the remaining question is how to weigh each edge in the projected network. Our final goal is to use personalized PageRank, the most well-known proximity measure, to score items based on the probability that a random surfer starting at the target user, will reach at each representative. Those probabilities can be adjusted by assigning weights to the edges of the graphs resulted from the projection process. To preserve the information available in original data, we will need to assign a weight to each link, $e_{ij}$ in a projection graph, based on the transition probability that the random surfer will reach from $i$ to $j$ in the original graph TPG when following meta-paths in $\theta_{alg}$. In other words, in a reasonable projection process, TPG must be projected to a weighted directed network with adjacency matrix $M$ where $M(i,j)$ approximates the probability of transition from $i$ to $j$ in the original network (TPG) when the transition paths are defined by a set of meta-paths $\theta$. Accordingly, we first estimate the transition matrix $T_\rho$ that reflects the transition probability from $i$ to $j$ using a meta-path $\rho \in \theta$. Then, we aggregate the whole set of transition matrices to infer adjacency matrix $M_\theta$ of the projected network over $\theta$.

To calculate the transition matrix of each meta-path, we define the basic adjacency matrices of the projected networks as $T_{up}, T_{pr}, T_{pu}$ and $T_{rp}$ as

$$T_{up}(i,j) = \begin{cases} \frac{1}{N(u_i)}, & (u_i, p_j) \in E_{UP} \\ 0, & o.w \end{cases} \quad (2)$$

$$T_{rp}(i,j) = \begin{cases} \frac{1}{N(r_i)}, & (p_j, r_i) \in E_{PR} \\ 0, & o.w \end{cases} \quad (3)$$

$$T_{pu}(i,j) = \begin{cases} \frac{1}{N(p_i)}, & (u_j, p_i) \in E_{UP} \\ 0, & o.w \end{cases} \quad (4)$$

$$T_{pr}(i,j) = \begin{cases} \frac{1}{N(p_i)}, & (p_i, r_j) \in E_{PR} \\ 0, & o.w \end{cases} \quad (5)$$

Where $N(v)$ indicates the number of adjacent nodes to a node v in TPG. $E_{UP}$ is the set of edges between users and preferences, and $E_{PR}$ is the set of edges between preferences and representatives.

In the next step, the transition matrix of each meta-path is obtained through $T_{v_1\ldots v_k} = \prod_{i=1}^{k-1} T_{v_i v_{i+1}}$ Where $v_1 \ldots v_k$ indicates a meta-path with length k starting from node type $v_1$, and, ending at node type $v_k$. For instance, the basic adjacency matrices of $T_{up}, T_{pr}$, should be multiplied to achieve the transition matrix that represents the probability of transition among nodes using a meta-path UPR or $T_{UPR} = T_{up} \times T_{pr}$.

So, we can define the adjacency matrix of the projected networks for the set of projection meta-paths (i.e $\theta$) through linear combination of adjacency matrices of its members using Eq. 6-8

$$M_{\theta_{UNC}}(i,j) = \begin{cases} \frac{T_{UPU}(i,j)}{W_{UPU}(i)+W_{UPR}(i)} & i,j \in U \\ \frac{T_{UPR}(i,j)}{W_{UPU}(i)+W_{UPR}(i)}, & i \in U, j \in R \\ \frac{T_{RPU}(i,j)}{W_{RPU}(i)}, & i \in R, j \in U \end{cases} \tag{6}$$

$$M_{\theta_{PNC}}(i,j) = \begin{cases} \frac{T_{UPU}(i,j)}{W_{UPU}(i)} & i \in U, j \in P \\ \frac{T_{UPR}(i,j)}{W_{UPR}(i)}, & i \in P, j \in R \end{cases} \tag{7}$$

$$M_{\theta_{RNC}}(i,j) = \begin{cases} \frac{T_{UPR}(i,j)}{W_{UPR}(i)}, i,j \in U \\ \frac{T_{RPU}(i,j)}{W_{RPU}(i)}, i,j \in R \end{cases} \tag{8}$$

Where $W_{UPU}(i) = \sum_{j=1}^{|U|} T_{UPU}(i,j)$, $W_{UPU}(i) = \sum_{j=1}^{|R|} T_{UPR}(i,j)$, and $W_{RPU}(i) = \sum_{j=1}^{|U|} T_{RPU}(i,j)$

## 6.2. Top-N recommendation through projected network

Once the networks constructed, we can use personalized PageRank on these networks to infer a recommendation. Personalized PageRank with a target user $u$ can be calculated by iteratively computing Eq.9

$$PPR_t^u = \alpha PPR_{t-1}^u M + (1-\alpha)d \tag{9}$$

where the $PPR_t^u$ denotes the personalized PageRank values of the target user $u$ and $d$ is the personalization vector that is a one-hot encoding for the target user $u$. We define the scoring function for each representative r, as $score(u,r) = PPR_\infty^u(r)$. Usually PPR-values converge in a small number of iteration, 20 iterations in our experiments. We can score different representatives through user-based, preference-based, and representative-based NC-Rank perspectives by substituting M with one of transition matrices $M_{\theta_{UNC}}$, $M_{\theta_{PNC}}$ or $M_{\theta_{RNC}}$. When score values of representatives are calculated, we can rank items through aggregating the score values of their corresponding representatives as in Eq.10.

$$rank(i) = score(u, i_d) - score(u, i_u) \tag{10}$$

## 7. Experimental Setting and results

We conducted a comprehensive set of experiments to evaluate the performance of ReGRank framework which models user-based, preference-based, and representative-based NB-Rank models. We denote the user-based variant of ReGRank as U-ReGRank, preference-based variant as P-ReGRank, and the representative-based variant as R-ReGRank.

We assess these algorithms on two publicly available datasets, Movielens100K, Movielens1M, and FilmTrust that are widely used for evaluating recommendation systems [1,3–5]. Both Movielens datasets are comprised of ratings in 5 levels (i. e. 1, 2, … , 5).while, FilmTrust contains rates from 1 to 4 at 8 levels (i. e. 1, 1.5, 2, … ,4).

We evaluate the recommendation algorithms through the well-known evaluation methodology in collaborative ranking literature[18,20,26,32], we split a fixed number UPL of ratings of each user as his training data, and his other ratings will be considered as test samples. This methodology is widely accepted in the community for several reasons: First, it enables us to compare the performance of algorithms under different sparsity. Second, the number of higher ratings in the test set simulates real world recommender systems that should only suggest a small number of items among a large number of unrated items[3,28]. For each value of UPL, 5 variants of training sets are generated via random sampling, and the average of the performance on their corresponding test sets is reported. In this paper, we assess the performance of algorithms with regards to data sparsity via changing UPL from 10 to 50 for all datasets. We should ensure that for each UPL, each user should have at least 10 items in the test as recommendation algorithms are evaluated through their Top-10 recommendation.

We assess the performance of algorithms using Normalized Cumulative Discounted Gain (NDCG) of their Top-N recommendation. Let $U$ be the set of users, $RL_u$ be the recommendation list for user $u$, and $IRL_u$ be the ideal recommendation list for user $u$. Then, we can calculate the average NDCG of Top-N using Eq.11

$$NDCG@topN = \frac{1}{|U|} \sum_{u=1}^{|U|} \frac{DCG_{RL_u}}{DCG_{IRL_u}} \tag{11}$$

Where $DCG_{RL_u}$ is computed through Eq. 12

$$DCG_{RL_u} = \sum_{i=1}^{topN} \frac{2^{rel_i} - 1}{log(i+1)} \tag{12}$$

Where rel_i is the relevancy of i-th item in the recommendation list. In addition to recommendation quality, we assess scalability of ReGRank framework through measuring its recommendation time.

**7.1. Baseline algorithms**

We assess the performance ReGRank against the state-of-the art neighbor-based collaborative ranking and graph-based recommendation algorithms. Also, we compare it to CofiRank, the most well-known matrix factorization algorithm for collaborative ranking. We briefly explain the algorithms below:

- **EigenRank:** EigenRank [12] is the most famous algorithm in the family of Neighbor-based collaborative ranking algorithm which uses Kendall correlation to find similarities between users' ranking, and then, it uses a random walk algorithm to aggregate the pairwise preferences and infer the total ranking of target user.
- **NN-GK:** NN-GK [5] is a recent neighborhood recommendation algorithm for pairwise preference dataset. NN-GK calculate users' similarity through Kruskal's gamma (GK) that is equivalent to Kendall correlation in rating datasets. Thereafter, it score each item through summing up the neighbors' opinions over its comparison with other items.
- **SibRank:** SibRank [18] is a state of the art neighbor-based collaborative ranking algorithm that calculates users' similarities based on signed multiplicative rank propagation from a target user's node on a signed bipartite preference network.
- **PrefRank:** PrefRank [7] presents another novel approach, which transforms users' pairwise preferences to a preference score and then, uses a traditional rating approach to calculate users and items' similarities. We call its user-based variant as U-PrefRank and its item-based variant as I-PrefRank.
- **GRank:** GRank [20] is the state of the art algorithm that processes TPG to make recommendations. GRank applies personalized PageRank on TPG in order to estimate the desirability/undesirability of items. Thereafter, to infer the overall score of an item, it aggregates the scores of their corresponding representative nodes.
- **CofiRank:** CofiRank[31] lies in one of the state-of-the-art algorithms in the class of matrix factorization collaborative ranking. CofiRank learns the latent factors of users and items while optimizing a structured loss function. The CofiRank's framework has been adapted for ordinal and NDCG ranking loss functions [32]. CofiRank-Ordinal minimizes the ordinal loss function that refers to the number of discordant preferences in the recommendation list On the other hand, CofiRank-NDCG triers to maximize the NDCG that reflects quality of recommendation list in subject to the ideal one. We used the publicly available code for CofiRank while setting the optimal parameter values that is provided in [32]

We implemented traditional neighbor-based collaborative ranking algorithms such as EigenRank, NN-GK, SibRank, and PrefRank using neighborhood size of 100 that is reportedly the neighborhood size that leads to the best performance for those algorithms. Additionally, we set the damping factor as 0.85 for GRank, EigenRank, SibRank, and different classes of ReGRank framework.

## 7.2. Performance analysis

We first analyze and compare the performance of different classes of ReGRank comparing to each other. As seen in Table 1-3, U-ReGRank and R-ReGRank significantly outperform P-ReGRank in lower values of UPL. For instance, in case of UPL=10, R-ReGRank and U-ReGRank achieve NDCG@10 of 71% for FilmTrust, 69% for ML100K, and 71% for ML-1M. On the other hand, P-ReGRank's performance does not exceed 69% in FilmTrust, 63% in ML100K, in 65% for ML1M when UPL=10. On the other hand, in higher UPL values, P-ReGRank performs similar to R-ReGRank and U-ReGRank in Movielens datasets, and even, 2% better in FilmTrust dataset. That result simply can be explained through the dataset's sparsity; in sparse datasets, users have rarely common pairwise preferences (e.g. Fig. 4b), and so the similarities between pairwise preferences are not reliable enough. In such a situation, type-2 similarity between users or similarity between representatives are more informative. On the other hand, as UPL increases, users have more common pairwise preferences, and consequently, similarity between pairwise preferences are more reliable for the recommendation. It should be noted that P-ReGRank could not achieve high performance in case of UPL=50 in FilmTrust dataset due to its high sparsity that is consequence of low number of users having more than 50 ratings in FilmTrust dataset. More clearly, in such a dataset, pairwise preferences' similarities are estimated based on a small number of users that can't provide enough information for an accurate estimation.

We also compare ReGRank algorithms to traditional neighborhood collaborative ranking algorithms and also GRank, the other state of the art recommendation approach over TPG. Experimental results showed the superiority of ReGRank over GRank in the majority of evaluation conditions especially when UPL takes low values. For instance, in case of UPL=10, the ReGRank shows an improvement of 2% in FilmTrust, 7% in Movielens100K, and 6% in Movielens1M. The reason is that in sparse datasets, TPG contains a large number of paths in form of RPR and PRP which are not reliable is any neighborhood recommendation approach. Therefore, GRank fails to make accurate recommendations in sparse datasets. In higher UPL values, users have more pairwise preferences that form a large number of reliable recommendation paths passing users and pairwise presences. Therefore, unreliable recommendation paths have a smaller effect on the scoring mechanism of GRank in higher values of UPL.

We also compare ReGRank algorithms to traditional NB-Rank algorithms, EigenRank, SibRank, and, PrefRank. ReGRank algorithms significantly outperform all other algorithms in majority of evaluation conditions. For instance, in UPL=30 in Movielens1M, ReGRank algorithms improve NDCG@10 up to 21% compared to PrefRank, up to 4% compared to EigenRank, NN-GK and SibRank. An interesting point is that the performance of different classes of PrefRank is significantly lower than other neighborhood algorithms which calculate users' similarity over pairwise preferences. This result implies that transforming users' rankings or pairwise preferences to numerical scores will probably eliminate some useful information and will decrease the recommendation's performance.

Finally, we compare our algorithm to different classes of CofiRank as representatives for matrix-factorization approach. ReGRank significantly outperforms both CofiRank-ordinal and CofiRank-NDCG in all experimental conditions. As an example, in case of UPL=50 in FilmTrust dataset, U-ReGRank, P-ReGRank, and R-ReGRank has gained NDCG@10 of 67%, 66%, and 67%, while, CofiRank-NDCG and CofiRank-Ordinal showed performance of 60% and 63%, respectively. An interesting point is that unlike neighborhood algorithm, CofiRank-NDCG lose its performance in higher UPL values. For instance, the performance of CofiRank-NDCG is 68% and 60% in movielense1m dataset when UPL=10 and UPL=50, respectively. Nevertheless, U-ReGRank increase its performance from 71% to 75% when UPL changes from 10 to 50 in Movielens1M. That is because neighborhood algorithms can capture the local taste of users and improve their recommendation accuracy in case of higher UPL value. Nevertheless, model-based algorithms typically learn the model that fits to the whole data. More clearly, model-based algorithms learn the global tastes of users and miss information that is available in local users' neighborhood.

Table 1. Comparison of algorithms in terms of NDCG@10 in Movielens100k dataset

| Algorithm | UPL=10 | UPL=20 | UPL=30 | UPL=40 | UPL=50 |
|---|---|---|---|---|---|
| **U-ReGRank** | 0.676±0.027 | 0.710±0.003 | **0.719±0.005** | **0.725±0.004** | 0.722±0.002 |
| **P-ReGRank** | 0.628±0.001 | 0.656±0.003 | 0.696±0.004 | 0.716±0.005 | **0.724±0.002** |
| **R-ReGRank** | **0.692±0.004** | **0.711±0.006** | 0.718±0.003 | 0.724±0.003 | 0.720±0.003 |
| **GRank** | 0.624±0.021 | 0.658±0.011 | 0.694±0.004 | 0.715±0.004 | 0.717±0.005 |
| **U-PrefRank** | 0.575±0.002 | 0.556±0.004 | 0.552±0.003 | 0.550±0.004 | 0.537±0.004 |
| **I-PrefRank** | 0.576±0.003 | 0.558±0.002 | 0.551±0.003 | 0.551±0.002 | 0.535±0.003 |
| **EigenRank** | 0.600±0.020 | 0.656±0.008 | 0.686±0.005 | 0.695±0.004 | 0.697±0.005 |
| **SibRank** | 0.648±0.008 | 0.671±0.004 | 0.692±0.006 | 0.705±0.005 | 0.710±0.004 |
| **NN-GK** | 0.594±0.002 | 0.658±0.012 | 0.694±0.004 | 0.715±0.004 | 0.717±0.005 |
| **CofiRank-NDCG** | 0.631±0.011 | 0.617±0.030 | 0.623±0.013 | 0.619±0.008 | 0.616±0.006 |
| **CofiRank-Ordinal** | 0.605±0.015 | 0.607±0.026 | 0.618±0.018 | 0.627±0.013 | 0.627±0.004 |

Table 2. Comparison of algorithms in terms of NDCG@10 in Movielens1M dataset

| Algorithm | UPL=10 | UPL=20 | UPL=30 | UPL=40 | UPL=50 |
|---|---|---|---|---|---|
| **U-ReGRank** | 0.718±0.018 | 0.739±0.022 | 0.745±0.019 | **0.750±0.018** | 0.753±0.017 |
| **P-ReGRank** | 0.652±0.006 | 0.695±0.006 | 0.729±0.019 | 0.746±0.022 | **0.755±0.023** |
| **R-ReGRank** | **0.723±0.020** | **0.739±0.021** | **0.744±0.018** | 0.748±0.017 | 0.741±0.016 |
| **GRank** | 0.655±0.019 | 0.694±0.013 | 0.728±0.002 | 0.744±0.010 | 0.752±0.014 |
| **U-PrefRank** | 0.550±0.003 | 0.541±0.004 | 0.535±0.005 | 0.528±0.005 | 0.527±0.005 |
| **I-PrefRank** | 0.550±0.004 | 0.542±0.006 | 0.536±0.007 | 0.529±0.007 | 0.527±0.006 |
| **EigenRank** | 0.614±0.004 | 0.699±0.001 | 0.703±0.009 | 0.701±0.012 | 0.692±0.012 |
| **SibRank** | 0.674±0.009 | 0.701±0.004 | 0.711±0.011 | 0.716±0.014 | 0.723±0.013 |
| **NN-GK** | 0.620±0.006 | 0.694±0.024 | 0.707±0.021 | 0.709±0.020 | 0.709±0.019 |
| **CofiRank-NDCG** | 0.685±0.002 | 0.685±0.024 | 0.670±0.016 | 0.657±0.003 | 0.645±0.003 |
| **CofiRank-Ordinal** | 0.637±0.017 | 0.646±0.009 | 0.663±0.001 | 0.669±0.002 | 0.675±0.008 |

Table 3. Comparison of algorithms in terms of NDCG@10 in FilmTrust dataset

| Algorithm | UPL=10 | UPL=20 | UPL=30 | UPL=40 | UPL=50 |
|---|---|---|---|---|---|
| U-ReGRank | 0.710±0.007 | 0.702±0.005 | 0.725±0.006 | 0.818±0.005 | **0.675±0.011** |
| P-ReGRank | 0.697±0.005 | **0.712±0.007** | 0.732±0.006 | 0.826±0.004 | 0.661±0.008 |
| R-ReGRank | **0.713±0.007** | 0.703±0.004 | 0.724±0.004 | 0.819±0.004 | **0.675±0.009** |
| GRank | 0.692±0.004 | 0.712±0.006 | **0.737±0.006** | **0.830±0.003** | 0.664±0.006 |
| U-PrefRank | 0.680±0.003 | 0.681±0.002 | 0.701±0.007 | 0.800±0.006 | 0.594±0.011 |
| I-PrefRank | 0.680±0.002 | 0.681±0.002 | 0.701±0.006 | 0.800±0.006 | 0.595±0.006 |
| EigenRank | 0.684±0.005 | 0.689±0.004 | 0.710±0.005 | 0.812±0.006 | 0.607±0.011 |
| SibRank | 0.683±0.005 | 0.688±0.007 | 0.710±0.004 | 0.814±0.005 | 0.612±0.014 |
| NN-GK | 0.692±0.005 | 0.705±0.002 | 0.726±0.013 | 0.820±0.004 | 0.654±0.013 |
| CofiRank-NDCG | 0.684±0.003 | 0.686±0.000 | 0.711±0.003 | 0.810±0.004 | 0.607±0.016 |
| CofiRank-Ordinal | 0.668±0.002 | 0.667±0.003 | 0.707±0.008 | 0.824±0.003 | 0.632±0.012 |

## 7.3. Scalability analysis

We also analyze the scalability of different neighborhood algorithms in terms of computational complexity and running times. Figure 9 illustrates the running of neighborhood algorithms measured in seconds on a Linux based PC running an Intel core i7-5820K processor at 3.3 GHz with 32GB of RAM.

Note that is not plausible to compare the running time of neighborhood approach and matrix factorization algorithms as neighborhood recommendation typically makes an up-to-date recommendation which requires online processing, while, model-based algorithms require an offline learning to learn latent factors of users and items. Here, we focus on the recommendation process in neighborhood recommendation approach.

Given $m$ as the number of users and $n$ as the number of items, projecting TPG over UPU path will maximally generate $m^2$ edges and projecting over UPU/UPR will maximally generate $2mn$ edges. Accordingly, $g_{UNC}$ contains $m^2 + 2mn$ links, $g_{PNC}$ contains $m^2 + mn$ links, and $g_{RNC}$ contains $2mn$ links. ReGRank make recommendation through personalized PageRank of target user over $g_{UNC}$, $g_{PNC}$, and $g_{RNC}$. Remind that personalized PageRank can be implemented at computational complexity of $O(te)$ where $e$ denotes the number of links and $t$ is the number of iterations required for convergence. Therefore, R-ReGRank makes recommendation to a target user at computational complexity of $O(tmn)$, while, P-ReGRank and U-RegRank would recommend at computational complexity of $O(tml + tm^2)$. So, R-ReGRank is more scalable than other variants of ReGRank. This is also reflected in their running time of different extensions of ReGRank framework; For instance, R-ReGRank makes recommendation in 0.07 seconds, P-ReGRank in 0.73 seconds, and U-ReGRank in 0.8 seconds in Movielens1M when UPL=50.

The running time of R-ReGRank is also up to 40 times less than U-PrefRank, 60 times less than I-PrefRank, 80 times less than pairwise-oriented neighborhood collaborative ranks that are NN-GK, SibRank, and EigenRank , and 180 times less than GRank in movielens1M dataset where UPL is set to 40. These running times are also compatible with their computational complexity that is $O(tmn + tn^2)$ for GRank, $O(mn^2 + kn)$ for I- PrefRank, $O(mn + kn)$ for U-PrefRank, $O(mn^2 + kn^2 + n^2)$ for NN-GK, SibRank, and EigenRank with neighborhood size of $k$.

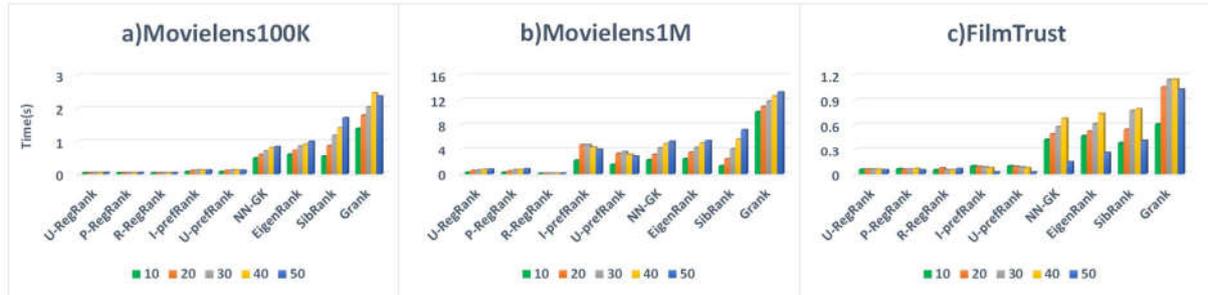

Fig. 9. Runtime of recommendation process in neighborhood collaborative ranking algorithms over Movielens1M dataset.

## 8. Conclusion

In this paper, we investigated the semantics behind different meta-paths of TPG, and their relations to neighborhood collaborative ranking algorithms. Thereafter, we present a framework, ReGRank, to make recommendation through meta-paths that are reliable from different perspective of neighborhood collaborative ranking. ReGRank models reliable recommendation meta-paths of NC-Rank algorithms, and then, projects TPG to other network containing these reliable recommendation meta-paths. Experimental results showed significant improvement of the suggested framework over well-known and state of the art graph-based and neighborhood collaborative ranking algorithms. Furthermore, we observed that using paths that are consistent with reliable user-based and representative-based meta-paths are more effective in sparse datasets, while, using paths that are in harmony with preference-based meta-paths leads to better performance in denser datasets.

Despite overall improvement of ReGRank framework over existing algorithms, there are several interesting directions for extending the current work. ReGRank scores items based on a diverse set of reliable meta-paths. Although those meta-paths may have different values in scoring items, ReGRank, in its current form, does not make any effort to weight them based on their importance. One possible direction for extending the current research is to assign such weights to the reliable meta-paths. Furthermore, ReGRank follows an off-line algorithm to weigh each edge in the projected networks and so it would be valuable to design an online algorithm for updating the weights of links as new data is inserted to the system. Finally, it is possible to extend the current research to support reliable graph-based recommendation using heterogeneous information networks containing other types of information like content and context.

## References

[1] I. AvSegal, Z. Katzir, K. Gal, EduRank : A Collaborative Filtering Approach to Personalization in E-learning, in: Educational Data Mining 2014, 2014: pp. 68–75.
[2] N. Chowdhury, X. Cai, C. Luo, BoostMF : Boosted Matrix Factorisation, in: Machine Learning and Knowledge Discovery in Databases (ECML/PKDD), 2015: pp. 3–18.
[3] K. Christakopoulou, A. Banerjee, Collaborative Ranking with a Push at the Top, in: WWW 2015: Proceedings of the 24th International Conference on World Wide Web, 2015: pp. 205–215.
[4] P. Han, B. Xie, F. Yang, R. Shen, A scalable P2P recommender system based on distributed collaborative filtering, Expert Systems with Applications. 27 (2004) 203–210.
[5] S. Kalloori, F. Ricci, M. Tkalcic, Pairwise Preferences Based Matrix Factorization and Nearest Neighbor Recommendation Techniques, Proceedings of the 10th ACM Conference on Recommender Systems - RecSys '16. (2016) 143–146.
[6] J.K. Kim, H.K. Kim, Y.H. Cho, A user-oriented contents recommendation system in peer-to-peer architecture, Expert Systems with Applications. 34 (2008) 300–312.
[7] J. Lee, D. Lee, Y.C. Lee, W.S. Hwang, S.W. Kim, Improving the accuracy of top-N recommendation using a preference model, Information Sciences. 348 (2016) 290–304.
[8] S. Lee, S. Park, M. Kahng, S. Lee, PathRank : Ranking nodes on a heterogeneous graph for flexible hybrid recommender systems, Expert Systems with Applications. 40 (2013) 684–697.
[9] S. Lee, O. Ridge, S. Lee, B. Park, PathMining : A Path-Based User Profiling Algorithm for Heterogeneous Graph-Based Recommender Systems, in: Proceedings of the Twenty-Eighth International Florida Artificial Intelligence Research Society Conference PathMining:, 2015: pp. 519–522.
[10] G. Li, Z. Zhang, L. Wang, Q. Chen, J. Pan, One-class collaborative filtering based on rating prediction and ranking prediction, Knowledge-Based Systems. 124 (2017) 46–54.
[11] J. Liu, C. Wu, Y. Xiong, W. Liu, List-wise probabilistic matrix factorization for recommendation, Information Sciences. 278 (2014) 434–447.
[12] N. Liu, Q. Yang, EigenRank : A Ranking-Oriented Approach to Collaborative Filtering, in: Proceedings of the 31st Annual International ACM SIGIR Conference on Research and Development in Information Retrieval ACM, 2008: pp. 83–90.
[13] N.N. Liu, Q. Yang, EigenRank: A Ranking-oriented Approach to Collaborative Filtering, Proceedings of the 31st Annual International ACM SIGIR Conference on Research and Development in Information Retrieval. (2008) 83–90.
[14] L. Meng, J. Li, H. Sun, WSRank: A Collaborative Ranking Approach for Web Service Selection, in: 2011 IEEE 11th International Conference on Computer and Information Technology, Ieee, 2011: pp. 103–108.
[15] L. Page, S. Brin, R. Motwani, T. Winograd, The PageRank citation ranking: bringing order to the web., (1999) 1–17.
[16] W. Pan, L. Chen, Cofiset: Collaborative filtering via learning pairwise preferences over item-sets,' Siam, Training. (2013) 180–188.
[17] W. Pan, H. Zhong, C. Xu, Z. Ming, Adaptive Bayesian personalized ranking for heterogeneous implicit feedbacks, Knowledge-Based Systems. 73 (2015) 173–180.
[18] B. Shams, S. Haratizadeh, SibRank: Signed bipartite network analysis for neighbor-based collaborative ranking, Physica A: Statistical Mechanics and Its Applications. 458 (2016) 364–377.
[19] B. Shams, S. Haratizadeh, Graph-based collaborative ranking, Expert Systems with

[20] B. Shams, S. Haratizadeh, Graph-based collaborative ranking, Expert Systems with Applications. 67 (2017) 59–70.
[21] C. Shi, B. Wang, P.S. Yu, HeteRecom : A Semantic-based Recommendation System in Heterogeneous Networks, Kdd. (2012) 1552–1555.
[22] C. Shi, P.S. Yu, B. Wu, Semantic Path based Personalized Recommendation on Weighted Heterogeneous Information Networks, Cikm. (2015) 453–462.
[23] Y. Shi, A. Karatzoglou, L. Baltrunas, Climf: learning to maximize reciprocal rank with collaborative less-is-more filtering, in: Proceedings of the Sixth ACM Conference on Recommender Systems ACM, 2012: pp. 139–146.
[24] Y. Shi, A. Karatzoglou, L. Baltrunas, M. Larson, xCLiMF : Optimizing Expected Reciprocal Rank for Data with Multiple Levels of Relevance, in: Proceedings of the 7th ACM Conference on Recommender Systems ACM, 2013: pp. 431–434.
[25] Y. Shi, M. Larson, A. Hanjalic, List-wise learning to rank with matrix factorization for collaborative filtering, in: Proceedings of the Fourth ACM Conference on Recommender System, 2010: pp. 269–272.
[26] Y. Shi, M. Larson, A. Hanjalic, Unifying rating-oriented and ranking-oriented collaborative filtering for improved recommendation, Information Sciences. 229 (2013) 29–39.
[27] Y. Sun, J. Han, X. Yan, P.S. Yu, PathSim : Meta Path-Based Top-K Similarity Search in Heterogeneous Information Networks, in: VLDB, 2011: pp. 992–1003.
[28] M.N. Volkovs, R.S. Zemel, Collaborative Ranking With 17 Parameters, in: Advances in Neural Information Processing Systems, 2012: pp. 2294–2302.
[29] S. Wang, S. Huang, T.-Y. Liu, J. Ma, Z. Chen, J. Veijalainen, Ranking-Oriented Collaborative Filtering: A Listwise Approach, ACM Trans Inf Syst. 35 (2016) 10:1--10:28.
[30] S. Wang, J. Sun, B.J. Gao, VSRank : A Novel Framework for Ranking-Based collaborative filtering, ACM Transactions on Intelligent Systems and Technology (TIST). 5 (2014) 51.
[31] M. Weimer, A. Karatzoglou, Maximum Margin Matrix Factorization for Collaborative Ranking, Advances in Neural Information Processing Systems. (2007) 1–8.
[32] M. Weimer, A. Karatzoglou, A. Smola, Improving maximum margin matrix factorization, Machine Learning. 72 (2008) 263–276.
[33] C. Yang, B. Wei, J. Wu, Y. Zhang, L. Zhang, CARES: a ranking-oriented CADAL recommender system, in: Proceedings of the 9th ACM/IEEE-CS Joint Conference on Digital Libraries ACM, 2009: pp. 203–212.
[34] Z.Y.Z. Yin, M. Gupta, T. Weninger, J.H.J. Han, A Unified Framework for Link Recommendation Using Random Walks, Advances in Social Networks Analysis and Mining (ASONAM), 2010 International Conference on. (2010).
[35] X. Yu, X. Ren, Y. Sun, Q. Gu, Personalized Entity Recommendation : A Heterogeneous Information Network Approach, in: 7th ACM International Conference on Web Search and Data Mining, 2014: pp. 283–292.